\documentclass[11pt]{article}
\def\itp#1{(\textit{#1}\/)}
\newtheorem{lemma}{Lemma}

\def\sqr{{\vcenter{\vbox{\hrule height.4pt%
	\hbox{\vrule width.4pt height5pt \kern5pt%
	\vrule width.4pt} \hrule height.4pt}}}}
\newcommand{\qed}{\hfill$\sqr$}
\newenvironment{proof}{\begin{trivlist}
\item[\hspace{\labelsep}{\bf\noindent Proof: }]
}{\hfill\qed\end{trivlist}}
\newsavebox{\boxtabbing}
{\end{tabbing}\end{minipage}\end{lrbox}%
\framebox[\columnwidth][l]{\usebox{\boxtabbing}}}
\newcommand{\doublespace}{\addtolength{\baselineskip}{.25\baselineskip}}
\newcommand{\remove}[1]{}
\def\height{\textit{height}}
\def\nextslot{\textit{nextslot}}

\def\leftchild{\textit{left}}
\def\rightchild{\textit{right}}
\def\val{\textit{val}}
\def\toggle{\textit{toggle}}
\def\root{\textit{root}}

\def\insertR{\texttt{insert}}
\def\deleteMin{\texttt{deleteMin}}

\def\verify{\texttt{verify}}

\def\findSlot{\texttt{findSlot}}

\def\upHeapify{\texttt{upHeapify}}
\def\balance{\texttt{balance}}
\def\deepLeaf{\texttt{deepLeaf}}
\def\downHeapify{\texttt{downHeapify}}
\def\nextPath{\texttt{nextPath}}

\def\leaf{\texttt{leaf}}
\def\leftFringe{\texttt{leftFringe}}
\def\swAncestor{\texttt{swAncestor}}
\begin{document}
\title{Available Stabilizing Heaps}
\author{Ted Herman\thanks{This work is supported
by NSF CAREER award CCR-9733541.} \\ 
University of Iowa \\ {\tt herman@cs.uiowa.edu} 
\and 
Toshimitsu Masuzawa \\ NAIST, Japan \\ {\tt masuzawa@is.aist-nara.ac.jp}} 
\date{12 July 2000}
\maketitle
\begin{abstract}
This paper describes a heap construction that supports 
insert and delete operations in arbitrary (possibly illegitimate)
states.  After any sequence of at most $O(m)$ heap
operations, the heap state is guaranteed to be 
legitimate, where $m$ is the initial number 
of items in the heap.  The response from each operation is
consistent with its effect on the data structure, 
even for illegitimate states.  The time complexity
of each operation is $O(\lg K)$ where $K$ is the capacity of the
data structure; when the heap's state is legitimate the  
time complexity is $O(\lg n)$ for $n$ equal to 
the number items in the heap.
\end{abstract}
\vspace*{-3ex}\par\noindent\textbf{Keywords:}  
data structures, fault tolerance, recovery, self-stabilization
\doublespace
\section{Introduction}

Increased visibility of systems emphasizes the 
theme of system availability.  When availability is not important,
systems may handle failures by stopping normal system activity
and restoring damaged data from a recent backup copy.  But stopping and 
restoring from a backup interferes with system availability, and in many 
instances it is preferable to let system services continue, even if 
the behavior of the services show temporary inconsistencies.  
Bookkeeping is intrinsic to system implementation, so data structures
are found at the core of system programs.  The objective of 
system availability motivates research of 
data structures with availability properties.  

(Self-) stabilization is the paradigm usually associated with recovery
from transient faults of unlimited scope \cite{D00}.  
Stabilization is traditionally associated with a distributed system, 
where each process can perpetually check and repair its variables.  
Our work departs from this traditional setting: we investigate a sequential, 
non-distributed data structure, supposing that the data structure 
is managed only by standard methods.  We further suppose that each 
method invocation starts cleanly (with no transient damage to internal
or control variables of that invocation), which resembles other 
work on fault tolerance \cite{ER90,LP90,FMRT96}.  The heap proposed
here also constrains operation behavior during the period of 
convergence to a legitimate state --- an issue not usually addressed 
by stabilization research (papers \cite{DH97,UKMF97,H00} are the exception).  

\section{Stabilizing Heap Construction} \label{implementation}

The data structure presented below is a variant of the standard 
binary heap \cite{CLR90} with a maximum capacity of $K$ items.
Two operations are defined for the heap, $\insertR(p)$, 
which inserts value $p$ into the heap, and \deleteMin(~), 
which returns and removes an item of least value from the heap.  
We say that an $\insertR(p)$ \emph{succeeds} if it inserts item $p$ 
into the heap and \emph{fails} otherwise.  The response to an 
$\insertR(p)$ invocation indicates success or failure by returning
``ack'' for success and ``heap full'' for failure.   
Similarly, a \deleteMin(~) \emph{succeeds} if it returns an item
and \emph{fails} by returning a ``heap empty'' indication.   

The heap construction described here is based on a binary, balanced
tree of $K$ nodes, rooted at a node named \root, and denoted by $A$.  
Each node $x$ in tree $A$ has two 
associated constants $x.\leftchild$ and $x.\rightchild$,
which refer respectively to the left and right child of $x$ in $A$. 
The symbol $\lambda$ denotes the  
absence a child:  $x.\leftchild=\lambda$ ($x.\rightchild=\lambda$) 
indicates that $x$ has no left (right) child in $A$.  
We suppose that $x.\leftchild$ and $x.\rightchild$ cannot 
be corrupted by a transient fault.  (A conventional
heap implementation by an array \cite{CLR90} satisfies this assumption, since 
there is a static mapping between parent and child.)  
Each node $x$ has three variable fields, $x.\val$, $x.\height$, and
$x.\nextslot$ used for storage and management of heap items in the tree. 

The field $x.\val$ may contain a heap item for node $x$.  
We use the symbol $\infty$, which is a value outside the
domain of possible heap items, to indicate the absence of 
a heap item at a particular node.  For convenience, let 
$\lambda.\val=\infty$ and let $y.\val\leq\infty$ hold by 
definition for all $y\in A$.  We define tree 
$T_A$ to be a truncation of tree $A$ that includes only 
nodes in a path of non-$\infty$ values.  
Formally, for any node $x$, let $x\in T_A$ iff 
$x.\val\neq\infty$ and either $x$ is \root\ 
or the child, with respect to $A$, 
of some $y$ satisfying $y\in T_A$ (the definition is recursive).
It follows that $T_A$ is empty if $\root.\val=\infty$.
A node $x\in T_A$ is a \emph{leaf} of $T_A$ iff 
$(x.\leftchild).\val=(x.\rightchild).\val=\infty$.
The expression $\langle T_A\rangle$ denotes the bag (multiset) 
given by $\{x.\val\;|\;x\in T_A\}$, and $|T_A|$ is 
the number of nodes in tree $T_A$.

Implementation of a binary heap should satisfy 
the balance property and the heap property.
The \emph{balance property} is that any heap of $m$ items is contained in 
a tree $T_A$ with height $O(\lg m)$.  The \emph{heap property} holds at
$x\in T_A$ iff $x.\val\leq y.\val$ for any $y$ child of $x$.  The 
data structure satisfies the \emph{heap property} iff the heap property
holds at every $x$, $x\in T_A$.   

\paragraph{Legitimate State.}
The fields of $A$'s nodes determine whether or not the 
tree is in a \emph{legitimate state}.  The state of $A$ is legitimate
iff \itp{i} for every $x\in T_A$, the heap property holds; 
\itp{ii} the balance property holds; 
\itp{iii} for every $x\in T_A$, $x.\height$ is the height of the 
subtree (in $T_A$) rooted at $x$; and \itp{iv} 
for every $x\in T_A$, $x.\nextslot$ equals the minimum distance from $x$
to a descendant $y\in T_A$ such that $y$ has fewer children in $T_A$ 
than $A$, and if no such descendant $y$ exists, $x.\nextslot$ can 
have any value satisfying $x.\nextslot\geq K$.

\paragraph{Basic Operations.}  Because conventional heap operations
are well known, we provide only sketches of the operation logic.  
Two internal routines \deepLeaf\ and \findSlot\ assist in node 
allocation and tree maintenance.   Let \deepLeaf(\root) be a recursive 
procedure that locates a node of $T_A$ having maximum depth: 
$\deepLeaf(x)$ compares the \height\ fields of $x$'s children,
and if one of them, say $y$, has greater height, then $\deepLeaf(x)$
returns $\deepLeaf(y)$;  and if both children have equal \height\ 
fields, then $\deepLeaf(x)$ returns $\deepLeaf(y)$ for some 
(possibly nondeterministic choice of) $y\in\{x.\leftchild,x.\rightchild\}$;
$\deepLeaf(x)$ returns $x$ if $x$ has no children.
A call to $\deepLeaf(\root)$
returns the symbol $\lambda$ if $T_A$ is empty. 

Let \findSlot(\root) be a recursive procedure locating 
a minimum-depth node $y$ in $A$ such that $y\not\in T_A$:
if $x$ has one child $y$ in $A$ such that $y\not\in T_A$, then
$\findSlot(x)$ returns $y$;  if $x$ has two children in $A$ such
that neither is in $T_A$, then $\findSlot(x)$ returns an arbitrary 
child of $x$; and  if $x$ has two children in $T_A$, $\findSlot(x)$ compares 
the \nextslot\ fields of $x$'s children and returns $\findSlot(y)$
for the child $y$ that has a smaller \nextslot\ field 
(if both children have equal \nextslot\ 
fields, then $y$ can be any child of $x$.)
An invocation of $\findSlot(\root)$ may fail to return 
an element $y$, and instead will return symbol $\lambda$, 
if no $y\not\in T_A$ can be located.  

The implementation of $\insertR(p)$ consists of assigning 
$y:=\findSlot(\root)$, and if $y=\lambda$, then $\insertR(p)$
responds with a ``heap full'' indication; 
otherwise, the following sequence executes.  First, the operation
assigns $y.\val,y.\height,y.\nextslot:=p,0,t$ where $t=0$ if $y$ 
has a child in $A$ and otherwise $t=K$.  Second,  
$z.\val:=\infty$ is assigned for every child $z$ of $y$ in $A$. 
Third, the operation calls $\upHeapify(y)$.  
The \upHeapify\ routine swaps values
of items on a path from $y$ to \root, until each parent 
has a value at most that of its child in the path.
As \upHeapify\ traverses the path from leaf to \root, it
also enforces, for each item $y$ on the path,
$y.\height:=1+\max((y.\leftchild).\height,(y.\rightchild).\height)$ 
and $y.\nextslot:=1+\min((y.\leftchild).\nextslot,(y.\rightchild).\nextslot)$
(with appropriate adjustments to these expressions
for cases of single or no children).

The implementation of $\deleteMin(~)$ consists of
$y:=\deepLeaf(\root)$, and if $y\neq\lambda$, saving 
$\root.\val$ for the response, then assigning
$\root.\val:=y.\val$ and $y.\val:=\infty$,    
and then calling $\downHeapify(\root)$.  Along the path from $y$ to \root, 
the \height\ and \nextslot\ fields are
also recomputed owing to the deletion of $y$.  
The \downHeapify\ routine swaps values as needed, 
along some path from \root\ to a leaf, so that the heap 
property is restored to $T_A$.  

\paragraph{Active Tree.}  If $A$ is not in a legitimate
state, it is still possible to consider the maximum fragment of $A$ 
that enjoys the heap property.  The \emph{active tree} $S_A$ is 
defined recursively by:  if $\root.\val=\infty$ then
$S_A$ is empty, otherwise $\root\in S_A$; and if $x\in S_A$ 
and $y$ is a child of $x$ (with respect to $A$)
such that $y.\val\neq\infty$ and
$y.\val\geq x.\val$, then $y\in S_A$. 

\paragraph{Operation Modifications for Stabilization.}  The definition of a 
legitimate state implies, for every $x\in S_A$ and $y\not\in S_A$
where $y$ is a child of $x$ in $A$, that the \val\ field of
$y$ is $\infty$.  In an illegitimate state, this condition need
not hold, though $S_A$ is defined even for illegitimate states.
Our first modification of operations (including internal 
routines such as \deepLeaf\ and \findSlot) is the following:
whenever an item $x\in S_A$ is encountered, it is first examined
to verify that every child $y$ satisfies $y.\val\geq x.\val$, 
and if this is not the case, then $y.\val:=\infty$ is immediately
assigned.  (The only exception to this modification are the heapify
routines, which are expected to encounter value reversals along
a particular path.)  The result of this modification is that 
operations consider children and leaves with respect to 
$S_A$ rather than $T_A$.  Observe that if a sequence of heap operations
could somehow encounter all the nodes of $T_A$, then $T_A=S_A$ would
hold as a result.  A second modification introduced below does this,
enforcing a scan of sufficiently many nodes of $T_A$ over any sequence
of heap operations so that $T_A=S_A$ will hold. 
We call this modification ``truncation'' since it
removes nodes from $T_A$ to enforce the heap property.  The truncation
modification is a convenience for our presentation --- another possibility
would be to treat $y.\val$ as equivalent to $\infty$ whenever 
$y.\val<x.\val$ for $x$ the parent of $y$, and adjusting the 
definition of legitimate state (and $T_A$) accordingly. 

The following lemma considers an operation applied to $A$
in an arbitrary (possibly illegitimate) state; for this lemma,
$T$ denotes $S_A$ prior to the operation and
$T'$ denotes $S_A$ after the operation. 

\begin{lemma} \label{avail}   
An operation applied to $A$ in an arbitrary state satisfies:
if \emph{$\insertR(p)$} succeeds, then 
$\langle T'\rangle=\langle T\rangle\cup\{p\}$;
if \emph{$\insertR(p)$} fails, then 
$\langle T'\rangle=\langle T\rangle$;
a \emph{\deleteMin} operations fails iff $T$ is empty;
if \emph{\deleteMin} returns value $q$, then 
$q=\min\langle T\rangle$ and
$\langle T'\rangle=\langle T\rangle\setminus\{q\}$;
and any operation completes in $O(\lg K)$ time.
\end{lemma}
\begin{proof}
Although \findSlot(\root) does not guarantee to find an available
position for an $\insertR(p)$ operation for illegitimate $A$, if 
\findSlot(\root) does return $r\neq\lambda$, then from the logic
of \findSlot, $r$ is a child of some node of $T$, and thus $T'$ 
will contain $p$ as a result of $\insertR(p)$. For a 
\deleteMin\ operation, $\deepLeaf(\root)$ returns some leaf of
$T$ (not necessarily at greatest depth in $T$) provided $T$ is 
nonempty, so \deleteMin\ returns \root.\val\ of $T$. 
The $O(\lg K)$ time bound is satisfied because any
path from \root\ to leaf in $T$ has length at most $\lg K$.
\end{proof}

The accuracy of \height\ and \nextslot\ fields is critical for 
maintaining the balance condition and locating an available node 
for heap insertion.  In an illegitimate state, these fields 
have arbitrary values.  Although heap operations recompute 
\height\ and \nextslot\ fields, such 
recomputation is limited to paths selected by the operations. 
The second change we make to operations is to add calls to 
a new routine \verify.  Each application of \verify\ works on
three objectives:  \itp{1} to apply truncation along one path $P$
from \root\ to a leaf of $S_A$, \itp{2} to assign \height\
and \nextslot\ fields from leaf to \root\ in $P$, and \itp{3}
to modify fields so that the next invocation of
\verify\ will select a path different from $P$.  
To support objective \itp{3},
we add a new binary field \toggle, with domain
$\{\ell,r\}$, to every node.  The path $P$ chosen by 
\verify\ is obtained by following \toggle\ directions
from \root\ until a leaf of $S_A$ is reached.  
Our intent is that $O(|S_A|)$ successive invocations of
\verify\ will visit all nodes of the active tree.

Figure \ref{verify} shows routine \verify.
Objectives \itp{1} and \itp{2} of \verify\ are achieved
with straightforward calculations.
Although not shown explicitly 
in the figure, \verify\ first checks values and assigns
$\infty$ if necessary to enforce the heap property, as 
needed for the truncation procedure.  The implementation
of objective \itp{3} is more complicated, using subordinate
routines \nextPath, \leftFringe, and \swAncestor.  
 
The first few lines of \verify\ in Figure \ref{verify} 
assign $x.\toggle$ for the case of $x$ having fewer than
two children:  in case $x$ has only a single child,
then $x.\toggle$ should be $r$ or $\ell$ according to
the location of its only child.  In case $x$ has no
children, $x.\toggle$ is assigned $\ell$.  The setup for a new
path occurs by call to $\nextPath(x)$, which only 
occurs when $x$ is a leaf of the active tree.  The idea of
$\nextPath(x)$ is to locate an ancestor $w$ of $x$ with 
$w.\toggle=\ell$ and change $w.\toggle$ to $r$, thereby 
setting up the ``next path'' for \verify\ to examine. 
The routine \leftFringe\ ensures that whenever such
a change of $w.\toggle$ from $\ell$ to $r$ takes place,
the \toggle\ fields are such that the leftmost path 
of the subtree rooted at $w.\rightchild$ will be selected for
this ``next path'' of \verify.  All lines but the last of
\leftFringe\ consider degenerate cases (single or
no children).  Not shown in the figure is the 
\swAncestor\ procedure: $\swAncestor(x)$ returns the nearest ancestor
$w$ of $x$ such that $w$ has two children in $S_A$ and $w.\toggle=\ell$;     
if no such ancestor $w$ exists, then $\swAncestor(x)$ returns $\lambda$.
Observe that if $\swAncestor$ does return $\lambda$, then 
$\nextPath(x)$ sets up the ``next path'' to be the leftmost
path starting from \root\ in $S_A$.  

\begin{lemma} \label{verscan}
If $\lfloor(|S_A|+1)/2\rfloor$ successive 
invocations of \emph{\verify(\root)}
are applied to arbitrary $A$, then as a result $T_A=S_A$ 
and properties \emph{\itp{i}}, 
\emph{\itp{iii}} and \emph{\itp{iv}} hold. 
\end{lemma}
\begin{proof}
Let $P$ denote the path of nodes examined via recursion for 
a given invocation of $\verify(\root)$.  
Observe that $\leftFringe(\root)$ is
called within this invocation of \verify\ iff $P$ is the rightmost
path within $S_A$ (otherwise \swAncestor\ would return a non-$\lambda$ value). 
By construction, $\leftFringe(x)$ for $x\neq\root$ sets up the leftmost
path in $S_A$ to the right of $P$.  Let $T_x$ be a subtree of $S_A$, 
rooted at $x$, with $m$ leaves.  If
$m$ successive \verify\ invocations examine the leaves of $T_x$, then 
\itp{i}, \itp{iii} and \itp{iv} hold for the nodes of $T_x$ afterwards
(this can be shown by induction on subtree height).  Let $S$ be a preorder
listing of the leaves of $S_A$;  $S$ has at most 
$\lfloor(|S_A|+1)/2\rfloor$ items since
any binary tree of $n$ items has at most $(n+1)/2$ leaves.  It is 
straightforward to show that any $|S|$ successive invocations of 
\verify(\root) visit the leaves of $S_A$ in an order corresponding to 
some rotation of sequence $S$. 
\end{proof}

The remaining modification to standard heap operations consists of
having each \insertR\ and \deleteMin\ operation begin 
with ``\verify(\root) ; \balance(\root)''.
A \balance(\root) invocation consists of deleting 
a leaf $r$ found by $\deepLeaf(\root)$ from the heap 
and then reinserting $r$ into the heap.  When all \height\ and 
\nextslot\ fields in the active tree have legitimate values, 
the effect of \balance(\root) is to move an item from a position
of maximum depth to a position of minimum depth.  Since 
these fields could be illegitimate, care must be taken in the 
implementation of \balance\ so that deleted leaf $r$ is, in any
case, reinserted into the heap (perhaps by reversing the delete
of $r$ if reinsertion fails).  
To see why the \balance(\root) call is needed, consider
an initial active tree with height $k$ that has only one leaf. 
Without the \balance(\root) call, if only \insertR\ operations 
are applied to the heap, then $O(2^k)$ operations 
would be required to bring the tree into balance.  


\begin{figure}[ht]
\framebox[0.86\columnwidth][l]{
\begin{minipage}{0.86\columnwidth}
\begin{tabbing}
x \= xxxx \= xx \= xx \= xx \= \kill
\> $\verify(x)$ \\
\>\> \texttt{if} $x.\leftchild=\lambda$ \texttt{then} $x.\toggle:=r$  \\
\>\> \texttt{if} $x.\rightchild=\lambda$ \texttt{then} $x.\toggle:=\ell$ \\
\>\> \texttt{if} $\neg\leaf(x)$ \texttt{then} \\
\>\>\> \texttt{if} $x.\toggle=r$ \texttt{then} $\verify(x.\rightchild)$ 
	\texttt{else} $\verify(x.\leftchild)$ \\
\>\> \texttt{else} $\nextPath(x)$ \\
\>\> \emph{calculate \& assign} $x.\height$, $x.\nextslot$ 
\end{tabbing} 
\hrule
\begin{tabbing}
x \= xxxx \= xx \= xx \= xx \= \kill
\> $\nextPath(x)$ \\
\>\> $w := \swAncestor(x)$ \\
\>\> \texttt{if} $w\neq\lambda$ \texttt{then}
	$w.\toggle:=r$ ; $\leftFringe(w.\rightchild)$ ; \texttt{return} \\
\>\> \texttt{else} $\leftFringe(\root)$
\end{tabbing}
\hrule
\begin{tabbing}
x \= xxxx \= xx \= xx \= xx \= \kill
\> $\leftFringe(x)$ \\
\>\> \texttt{if} $x=\lambda \;\vee\; \leaf(x)$ \texttt{return} \\
\>\> \texttt{if} $x.\leftchild=\lambda$ \texttt{then} $x.\toggle:=r$ 
		; $\leftFringe(x.\rightchild)$ ; \texttt{return} \\
\>\> \texttt{if} $x.\rightchild=\lambda$ \texttt{then} $x.\toggle:=\ell$ 
		; $\leftFringe(x.\leftchild)$ ; \texttt{return} \\
\>\> $x.\toggle:=\ell$ ; $\leftFringe(x.\leftchild)$ ; \texttt{return}
\end{tabbing}
\end{minipage}
}
\caption{$\verify(x)$ and subordinate procedures.}
\label{verify}
\end{figure}

\begin{lemma} \label{metrics}
Let $m=|S_A|$ for an arbitrary initial state of $A$.  
After any sequence of at most $m+1$ operations, 
$A$ satisfies properties \emph{\itp{i}}, 
\emph{\itp{iii}} and \emph{\itp{iv}} at all subsequent states. 
\end{lemma}
\begin{proof}
Each \insertR\ and \deleteMin\ operation invokes 
\verify(\root), however such operations also change the active
tree.  With respect to the sequence of \verify(\root) calls starting
from the initial state, each change to the active tree either occurs
in a subtree previously visited by \verify\ or occurs in a subtree
not yet visited by \verify.  In the former case, the tree modification
satisfies \itp{i}, \itp{iii} and \itp{iv} along the path from \root\ 
to the modified nodes.  In the latter case, a future \verify\ establishes
the desired properties.  After $d$ operations, the active tree
has at most $(m+d+1)/2$ leaves;  since each operation invokes 
\verify(\root), $d$ operations visit all leaves provided
$(m+d+1)/2\leq d$, by an argument similar to the proof of 
Lemma \ref{verscan}.  Therefore $m+1\leq d$ suffices.   
\end{proof}

\begin{lemma} \label{converge}
Let $m=|S_A|$ for an arbitrary initial state of $A$.  
After at most $O(m)$ heap operations, 
all subsequent states of $A$ are legitimate.
\end{lemma}
\begin{proof}
Lemma \ref{metrics} establishes that properties \itp{i}, 
\itp{iii} and \itp{iv} hold after $O(m)$ operations.  
In the rest of this proof, we assume that properties \itp{i}, 
\itp{iii} and \itp{iv} hold.  For the sake of generality
we suppose that $A$ is only loosely balanced:  assume there
exist constants $a$ and $b$ so that for any $t$, $0<t\leq K$, 
the minimum height $h_t$ taken over all subtrees of $A$ with $t$ nodes
that contain node \root\ satisfies $h_t\leq a+b\lg t$.  From 
this assumption, it follows that any subtree $T$ of $A$
rooted at \root\ with height exceeding $h_t$ is nonoptimal; 
furthermore, it follows that there is some node $w\in A$ not contained in
$T$, that is a child of some node of $T$, so that $w$ has depth
at most $h_t$.  

We define $gap(\alpha)$ for any state $\alpha$ to be a variant function.
\[ gap(\alpha)=\sum_{x\in S_A} v(x), ~ \textrm{where} ~ 
v(x) = \left\{ \begin{array}{ll} 1 & \textrm{if} ~
depth(x)>h_{|S_A|} \\ 0 & \textrm{otherwise} \end{array} \right. \] 
It is straightforward to show that once $gap$ is zero, any subsequent
operation application results in zero $gap$, and that zero $gap$ implies
balance.  If the initial $gap$ is some value $g>0$, any new item
inserted into the heap is placed at minimum depth, any \deleteMin\ 
removes a node at maximum depth,  
so $gap$ does not increase by the insert or delete operations.  
Moreover, every operation invokes \balance(\root), 
which decreases positive $gap$ by at least one, so within $g=O(m)$ 
operations, property \itp{ii} is established.
\end{proof}

\section{Availability and Stabilization}

The previous section presents a heap construction that 
is stabilizing (Lemma \ref{converge}) and also satisfies
certain properties expected of operations even when the 
data structure's state is illegitimate (Lemma \ref{avail}).
Lemmas \ref{avail} and \ref{converge} depend on the definition
of $S_A$.  Is there a characterization of availability and
stabilization independent of implementation specifics such
as $S_A$?  Such a characterization could be adapted to 
specify availability and stabilization for general types and
implementations of data structures. 

Let $\cal H$ be an infinite history of operations on a heap, 
that is, $\cal H$ is a sequence of \insertR\ and \deleteMin\ invocations and
corresponding responses.  We characterize a heap implementation in 
terms of properties of all possible operation histories, first
for the case of an initially empty heap.  Let $t$ denote a point
either before any operation or between operations in $\cal H$.  
If $t$ is before any operation, define $C_t = \emptyset$,  
otherwise let $C_t = I_t \setminus D_t$, where $I_t$ is the bag of
items successfully inserted prior to point $t$, and $D_t$ is
the bag of items returned by \deleteMin\ operations prior
to point $t$ (recall that success or failure of an operation is
judged by the response it returns). 
We call $C_t$ the \emph{heap content} at point $t$.  
Heap operations satisfy the following constraints:
\itp{a} a \deleteMin\ operation immediately following any point
$t$ fails iff $C_t = \emptyset$, and otherwise returns
$\min(C_t)$; \itp{b} an \insertR\ operation   
immediately following any point $t$ fails iff $|C_t|=K$, 
and otherwise returns ``ack''; \itp{c} 
the running time of any operation immediately following $t$ 
is $O(\lg |C_t|)$.  From \itp{a}--\itp{c} one can show the usual
heap properties, for instance, no \deleteMin\ returns an item not
previously inserted. 

The above characterization of heap behavior by history 
depends on $C_t=\emptyset$ for the initial point $t$. 
Availability is a relaxation of this characterization to 
allow arbitrary initial heap content.
Let $\cal P$ denote a history fragment starting
from an initially empty heap, that consists of 
at most $K$ successful \insertR\ operations.  
To specify behavior of $\cal H$ for an arbitrary initial heap, 
let ${\cal H}' = {\cal P}\circ{\cal H}$ (where $\circ$ denotes catenation). 
A heap implementation is \emph{available} if, for each history
$\cal H$ of operations, there exists $\cal P$ such that
${\cal H}'$ satisfies constraint \itp{a}, each operation in ${\cal H}'$ 
has $O(\lg K)$ running time, and any \insertR\ operation 
following any point $t$ fails
if $|C_t|=K$ (but is allowed to fail even if $|C_t|\neq K$).   
The construction of Section \ref{implementation} satisfies availability,
as shown by Lemma \ref{avail}, by
choosing $P$ to be a sequence of \insertR\ operations
for the items of the active tree at the initial state.  
The simplest heap implementation satisfying availability 
is one that returns a failing response to every operation 
($\cal P$ is empty and the heap content is continuously empty in
this case).

Stabilization is also a weakening of \itp{a}--\itp{c}.
Let ${\cal H}_t$ denote the suffix of history $\cal H$
following point $t$.  A heap implementation is 
\emph{stabilizing} if, for each history $\cal H$ of 
operations, there exists $\cal P$ (a history fragment of
successful \insertR\ operations) and a point $t$ such
that \itp{a}--\itp{c} hold for ${\cal P}\circ{\cal H}_t$.  
We call the history prefix ${\cal C}$ satisfying 
${\cal H}={\cal C}\circ{\cal H}_t$ the \emph{convergence
period} of $\cal H$.  The construction of Section \ref{implementation} is 
stabilizing by choosing $\cal P$ to contain \insertR\ 
operations for the active tree at some point $t$ that exists 
by Lemma \ref{converge}.  The definition of stabilization 
permits operations to arbitrarily succeed or fail 
during the convergence period, and a \deleteMin\ operation
could return a value unrelated to heap content within the
convergence period.  A plausible stabilizing heap implementation is one that
resets the heap content to be empty whenever some 
inconsistency is detected during the processing of an
operation (resetting the heap amounts to establishing
a legitimate ``initial'' state for subsequent operations). 

\section{Discussion}

The heap construction presented here satisfies desired availability
properties:  success or failure in an operation response is 
a reliable indication of the operation's result on the data structure.
We have not addressed the issue of relating heap damage to the 
extent of a fault --- if a fault somehow sets $\root.\val=\infty$
then the entire heap contents are lost by our construction.  Our 
intent is to separate concerns by first developing a stabilizing heap,
and then later adding logic for limited cases of corrupted items.
This is a topic for future work.

\par
\emph{Acknowledgment.}  We thank anonymous referees for helpful
comments.  We are especially grateful for careful criticisms by
Professor Mohamed Gouda, which guided us to improve the presentation.

\addtolength{\baselineskip}{-.25\baselineskip}
\end{document}